\documentclass[10pt]{article}
\usepackage[left=2cm,right=2cm,top=2cm,bottom=2cm,footskip=1cm]{geometry}
\usepackage[utf8]{inputenc}
\usepackage[style=apa,autocite=inline, doi=false,url=false,hyperref=false,uniquename=false]{biblatex}
\usepackage{endnotes}
\usepackage{array}
\usepackage{url}
\usepackage{kantlipsum}
\let\footnote=\endnote
\let\cite=\autocite

\usepackage[bottom]{footmisc}

\usepackage{hyphenat}
\usepackage{adjustbox}
\usepackage{graphicx}
\usepackage{float}
\usepackage{longtable}
\usepackage{setspace}
\usepackage{enumitem}
\usepackage{dirtytalk}
\usepackage{parskip}
\usepackage{csquotes}
\usepackage{courier}
\usepackage{spverbatim}
\title{The Psychogeography of Imaginary Places}
\date{}
\author{
  Michael Heron\\
  Chalmers University of Technology\\ 
  University of Gothenburg\\ 
  \texttt{heronm@chalmers.se}
  \and
  Pauline Belford \\
  Chalmers University of Technology\\ 
  University of Gothenburg\\ 
  \texttt{pauline.belford@chalmers.se}
  \and
  Klara Aune \\
  Chalmers University of Technology\\ 
  University of Gothenburg\\ 
  \texttt{klaraau@chalmers.se}}

\begin{document}
\twocolumn

\maketitle

\section{Abstract}
Psychogeography—the study of how environments shape emotion and behaviour—has long concerned itself with the emotional resonance of the physical, often through the idea of the dérive through the city.  Its philosophical core though is primarily in identifying affective relationships between the personal and the environmental, and that does not require the constraint of concrete.  

This paper extends psychogeographical practice into the realm of the imaginary, proposing a psychogeography of virtual and fictive spaces. Drawing on literary, Situationist, and contemporary psychogeographical traditions, we examine how the dérive might operate within the elastic spatiality and temporalities of video game worlds. We argue that digital environments, being wholly constructed, invite new forms of meaning-making and self-reflection. Through this reframing, games become both laboratory and landscape for a revitalised psychogeography—one attuned not only to the spirits of streets and cities, but also to the ghosts that haunt code, pixels, and play.

\section{Introduction}

Pyschogeography - a reflective discipline which exists within the liminal space between geography and psychology - is difficult to define in a rigorous manner.  The term refers to an amorphous grouping of variably intellectually-compatible practices, employed across a range of forms and formats, to a wide range of diverse goals and with equally diverse outputs.  It is a miasma of contradictory practices and incompatible theory, tied together primarily by loose conventions.   More than anything else, it can be argued, psychogeography is driven by a form of idiosyncratic curiosity that seeks to extract meaning from how we, as people, exist within and without the external environments in which we find ourselves.  As the psychogeographer Hou Je Bek has put it \cite{o2021psychogeography}:

    \say{Psychogeography is the fact that you have an opinion about a space the moment you step into it.}  

Or as Elkin puts it \cite{elkin2017flaneuse}:

    \say{...every neighbourhood has its own specific feeling, even if, superficially, they can all look alike.  You know when you step out of your neighbourhood.  You can feel it.}

Psychogeography has philosophical outcroppings within a range of disconnected disciplines, from literary theory and practice \cite {coverley2018psychogeography, tso2020literary} to urban design \cite{ellard2015places, mudie2016reading, mcintyre2013physical} to citizen archaeology \cite{sidaway2022psychogeography} from social theory  \cite {bridger2012psychogeography, bonnett2009dilemmas} to political revolution \cite {mason2023interventions} and to mysticism and the occult \cite {bonnett2017enchanted, smith2010contemporary}.   However, one area in which its application is notably lacking is in a domain that seems eminently applicable to its goals and processes - that of the experiences one may have within virtual environments, and imaginary places.  There is a solid core of literary tradition associated with psychogeography - the works of Baudelaire, Ballard, Poe, and more are often included within the loosely-defined canon of psychogeographical literature.  However, the inclusion of such works is almost always focused on the construction, rather than interpretation of the texts.  Psychogeography, as an anarchic discipline, includes William Blake and Thomas De Quincey as part of its intellectual heritage - but primarily because of how the authors engage with the practice rather than how their readers engage with the books.  The role of psychogeography in helping illuminate and describe our engagement with the imagination of other people remains an under-explored topic.

In this paper, the authors address this knowledge gap by expanding traditional models of psychogeography to include the fantastical and the ephemeral.  In this we seek to include psychogeographical practice as a feature of critical interpretation generally, but we seek to focus primarily on a medium where psychogeography is both eminently applicable and under-explored.  The worlds of video games are rich and resplendent with promise - the mindful exploration of the relationship between player and world can yield powerful insights that are both important for self-reflection and also transferable as a form of 'best practice' within the process of critical interpretation.  However, the common expectations and assumptions of psychogeography do not necessarily survive being transplanted from the physical world - even those eerier parts which resonate so forcefully with those of a spiritual bent.  The conventions of time and space which reality imposes upon us are only somewhat applicable within an imaginary world.  With a video game, the rules of movement, presence, and chronology may be played and experimented with, or simply ignored as a consequence of implementation complexity.   This does not invalidate the practice of psychogeography, in our view.  It does create a unique texture and set of considerations that must be properly explored and evaluated before the approach can be profitably employed within fictive worlds.

\section*{Psychogeography in Theory and in Practice}

The roots of psychogeography are deep within history, with its earliest traces encoded into archaeological records and primitive practice.  An emotional resonance between someone and their environment is a theme as old as humanity itself.   The earliest formal traditions of psychogeography are anchored in the literary - Defoe, Poe, Baudelaire, Blake and others.   Virgina Woolf wrote of the practice as 'street haunting' \cite{woolf2022street}, reflecting in the rambling prose of her essay the digressive nature of psychogeography itself.  The cliche of shows such as Sex and the City is that 'New York is actually the main character', and it is this ethos that is to be found in the literary record.  The works of these authors often have an explicit focus on the environment that is distinct from many of their contemporaries - an approach that blended a narrative of identity that brought together the nature of the people with the nature of their land.  Blake's work contrasted the 'dark satanic mills' against his conception of England as a form of New Jerusalem, charging the people of England with the task of its construction.  However, it is poems such as London \cite{wolfreys1998writing} in which we see the stirrings of the psychogeographer in the poet.  Brief though the poem is, it acts as an inventory of the character of the city and the indignities it inflicts on its inhabitants.  'But most thro' midnight streets I hear, how the youthful harlots curse, blasts the new-born infants ear, and blights with plagues the marriage hearse'.   

The city of Blake's London is oppressive, and the people within it afflicted by 'mind forged manacles'.   Wordsworth, with his famous wandering lonely as a cloud, is also a poet who has been adopted by the modern psychogeography movement \cite{sidaway2022psychogeography}.  His poem, Daffodils is a reflection on the moment, and how an environment can speak to its explorer in a language made up purely of sensory experience.  Greenock  in comparison echoes some of the sentiments of Blake with regards to a people influenced beyond tolerance by their surroundings.  'These crowded streets resound no plaintive ditty, as from the hive where bees in summer dwell'.   More explicitly we can see the autobiographical works of Thomas de Quincey, in Confessions of an English Opium Eater \cite{de1863confessions}, are frank about the relationship of addiction and his rambling through a city barely navigable through the fog of his own kaleidoscopic perceptions.  Edgar Allen Poe was similarly explicit in the way he wrote the relationships between the people and their environments.   Jean Rhys \cite{munroe2015haunted} encapsulated Svetlana Boym's conception of the city as 'an ideal crossroads between longing and estrangement, memory and freedom, nostalgia and modernity' \cite{boym2008future}.  These serve in many ways as the spiritual core of what psychogeography would become in the modern era. 

One might however begin to see the first stirrings of a psychogeographical \textbf{vocabulary} in the fifties.  A vocabulary distinct and differentiated from mystical, spiritual, literary, religious or intuitive framings.  This can be seen emerging through the works of  Guy Debord and the wider Situationist movement that he headed.  Debord defined psychogeography as 'The study of the precise laws and specific effects of the geographical environment, consciously organized or not, on the emotions and behaviors of individuals' \cite{debord1955introduction}.   In the hands of Debord and his contemporaries, psychogeography was a tool for social change - a mechanism for diagnosing and manipulating the emotional tenor of an environment to bring about ends more compatible with the goals of the Situationist movement.  It was explicitly political \cite{trier2019guy} - a tool for diagnosis of maladies so that they could be corrected, coincidentally, by the social solutions he and his allies advocated.  

The primary tool of Debord was what was known as the dérive - an unplanned, subconscious exploration of an environment separate and distinct from any specific goal or end-point.   Debord describes the dérive as 'an experimental mode of  behaviour linked to the conditions of urban society: a technique for hastily passing through varied environments'.  Debord encouraged the practice as a small-group activity \cite {debord1958theory}  where like-minded individuals with similar levels of psychogeographical experience could pool their observations.  The intention was of arriving at objective conclusions regarding the psychological contours of physical space \cite{debord1955introduction}.  Within this framing, Debord clearly saw the dérive as something akin to a scientific tool of accurate measurement.  Psychogeographers here are not serene explorers, or passive recipients of generalised insights - but are instead engaged in a hybrid quantitative-qualitative practice exploring hypotheses regarding emotional disorientation.  This latter trait being one of the pre-requisites for the emergence of the Situations for which the movement was named.  

Contemporary psychogeography is perhaps most often applied within an urban context, and it is here we can see where the lofty goals of Debord and his contemporaries begin to break down.  Despite the facade of objectivity implied by a Debordian dérive, the reality of lived experience is such that the psychological tenor of a location is too varied, too contextual, and too fragile to be something that can be sampled in the same way as one might dip a pipette into a container of river water.   The complexion of a physical place, particularly one in which other people are active, has too many facets to be captured with a tool as whimsical as a dérive.  Consider for example a residential street after a spate of unsolved burglaries, versus that same street on the day the school holidays end, versus that street on a warm summer's day.   A city, town, village, street, or even an individual home is a mosaic of geographically and chronologically overlapping lives and narratives.  Thousands of people are bit players in millions of stories - billions of individual nodes, knotted together in ways that defy extrication.  And within a dérive, particularly the group model osed by Debord, observation becomes adjustment.  It is hard to imagine that a bunch of middle aged academics wandering into a suburban cul-de-sac with notebooks and pens are going to experience anything truly authentic.   Connections to an environment are not a road of one-way traffic - the 'graph' in psychogeographer, as in geographer itself, is the Latin word for 'write'.  We write our psychologies on our environments as clearly as our environments write their resonance upon us.  To dérive through an area with the intention of understanding its emotional characteristics is to grab fistfuls of air out of an ever-changing sky.  

Within the original goals of psychogeography are specifically revolutionary aims - that the dérive's role as a diagnostic tool was to find those liminal spaces where change must be enacted.  Along with this was an explicit rejection of norms of ownership and territorialism - a rejection of the constraints as to where someone may wander.  However, in this is an awareness of the fact the physical world may be inconvenient without malice - that a mountain or a river was not put in our way by a vengeful god seeking to vex us.   Instead, we react to the realities of a cold, unfeeling universe without it being necessary to take any of it personally.  Even when thinking of human annotations of the environment we cannot necessarily read malice into location, or at the very least we must temper social considerations with physical ones.  The placement of a harbour by a body of water may reflect the politics of the decision making - why here, not there? - but there is also an almost anthropic principle at play.  The harbour exists by the water, because harbours cannot functionally exist elsewhere.  Unpicking how we might feel about our environments requires a careful process of examination of proximal causes, historical context, and real-world limitations.  This all through a lens which is explicitly - and intentionally - subjective.        

This Debordian pseudo-scientific conception of psychogeography has largely fallen out of favour with the modern movement.  Modern practitioners such as Sinclair \cite{tso2020literary, downing2016probably} ; Self \cite{matthews2016dissolving, self2007psychogeography} Ackroyd \cite{garayeva2021peter, tso2020literary}; Home \cite{home2016mind, collier2017our}; and more have all embraced a more personal and subjective form of psychogeography.  One in which the dérive is a tool for exploring the intersection of geography, history and the personal resonance of places.   In modern expression, the technique is not about the construction and exploration of a scientific hypothesis - it instead focuses on invoking the pleasure of the moment.   It is not a process of building evidence, rather one of meaning making. 

As a consequence of this approach - an appreciation of embodiment within social constraints - we also see the trajectory of the modern psychogeographical movement traced out in an arc that is almost despairing.  To engage with psychogeography within the urban environment is to enter into an ongoing seance where the only contactable spirits are the manifestation of banalisation \cite{coverley2018psychogeography, kapp2024tendency},  a globalised malaise that renders all new experiences as simply remixes.  All cities invariably converge towards being eerie doppelgangers, where the facades may differ and the semiotics may retain their originality, but at the core everything feels the same. 

 The commodification of physical space in urban environments has driven many psychogeographers to the peripheries - to explore the liminal spaces between urban and rural environments \cite {farley2012edgelands}, or to engage with the suburban wreckage of frustrated late-stage Capitalism \cite {papadimitriou2012scarp}.  Others, such as Self \cite{ self2007psychogeography} , have expanded the definition of psychogeography to reject the conception of a dérive as one that is linked to a geographical context.  For Self, the spark of consciousness that represents himself is the through-line that defines physicality.  In panel discussions he has described his approach as a form of 'jump-cut psychogeography'.  He describes the experience of walking to Heathrow Airport being intercut with the liminal experience of airplane travel, to then reconvene - psychologically uninterrupted - in another airport in another country.  Other psychogeographers seek out abandoned places or the 'edgelands' defined by Farley and Roberts as 'those familiar yet ignored spaces which are neither city nor countryside'.   The practice of psychogeography, within the physical world, can be conducted in many ways.  This is perhaps why the outputs of the process are so varied, ranging from sketches, tone poems, heat-maps, stories, character vignettes and musical pieces.  The forms in which codified insight take are as varied as the environments that psychogeographers wander through.

The city though remains the primary canvas for modern psychogeographers, and in this two cities in particular are inextricably linked to the practice - Paris, and London.  Paris, for the fact that some of the most prominent psychogeographers - in the form of the Baudellarian Flaneur or or the feminized Flaneuse \cite{elkin2017flaneuse}  - are forever tied into that context.  London for the fact its bones bury more deeply into the landscape than perhaps any industrial city in history.   What Paris and London have though, other than simple name recognition and a pre-existing connection to psychogeography - is an architecture that lends itself easily to the practice of the dérive.  The roads wind and wend.  There are abundant tributaries and avenues.  It is possible to get lost within, and between, the layout.  To explore London without local knowledge or a map is to find yourself adrift, permitting a kind of unselfconscious engagement within virgin environments.  Paris too is a city that is both urban powerhouse and geographical labyrinth.   One prominent psychogeographer once gave a beginner's guide to psychogeography - take a cup, take a map, put the cup down on a random page and draw a circle around its rim.  The circle, unconstrained by intention or geography, is the path you are to follow.   To accomplish this within Paris or London is to engage in a puzzle-solving exercise of examining stairways and dark, wending passages.   To do the same in a modern American city such as New York is to find yourself ironically frustrated by the systems of blocks and avenues that make the city so easily navigable.   \citeauthor{aarseth1997cybertext}  (\citeyear{aarseth1997cybertext}) coined the term Ergodic Literature to describe texts that require a non-trivial amount of effort to traverse.  One notable example of which is the book House of Leaves \cite{hansen2004digital}  which has stories within stories within stories along with text that is in code or reversed so it must be read in a mirror.  London and Paris, in this context, represent a form of Ergodic Traversal, in which it is a non-trivial task to get around.  New York, by contrast, offers the psychogeographer fewer opportunities to leave the beaten path for unpredictable outcomes.  

Cities though represent a primary destination for psychogeographers because they represent a concrete, as it were, accretion of humanity's thoughts and plans.  The smallest decision in the 15th century may lead to lasting urban-planning consequences in the 21st.  Within the wilderness, the architect of decision making is only occasionally humanity.  The brush of natural forces has a much bigger impact on the canvas of the landscape.   As such, it is difficult to gain a psychological appreciation of the simple scale of the planet because it is anchored within incomprehensible seas.  The author Timothy Clark has used the term 'Derangement of Scale' to highlight our inability to grasp the incompatibilities between human conception, cartographical scale, and the sheer enormity of the natural world \cite{clark2012derangements}.  The physical annotations of a city - the quirk of an alleyway or the odd shape of a building designed to fit within unconventional constraints - offer us chronological and physical anchor points that prevent such derangement even if they don't necessarily help build an accurate context.  In this respect, cities are a place for those of us in the present to commune with the conventions of the past.  

With all this it is perhaps no surprise that the psychogeographer often has a touch of the occult about them.  The links between spiritualism, psychogeography and urban exploration are deep and intricate encompassing everything from ghosts \cite{johnston2016walking}  to ley lines \cite{hanson2007mind}.  The psychogeographer is part geographer, part psychologist, but also part medium.  To explore environments for their emotional resonance is to engage in a dialogue with the ghosts of times gone by.  It is the act of  conducting a seance through mindful communion with the physical world.   Some of the most vocal participants in psychogeography also have one foot in the distinct but related discipline of hauntology.  The authors here would distinguish between psychogeography as a form of 'vibes in space' versus hauntology as 'vibes in time'.   Magic has often been argued to exist most strongly in the liminal spaces and places - between life and death, between sleep and wakefulness, and between the city and the wilderness.  In ancient times, a faerie circle marked the boundaries between this world and another.  Liminal spaces often have the quality of being eerie - places where our normal faculties and expectations are changed, or arrested.  The timelessness of a waiting room, or the heightened sense of threat in an underpass.    In the wilderness, our senses are attuned in one way - to nature, and to the forces underpinning it.  While genuinely wild spaces are increasingly rare on a crowded planet, natural spaces feel as if they are haunted not so much by spirits but rather by the absence of ghosts.  Cities, on the other hand, are constantly haunted.  All hauntings are real, as argued by \citeauthor{fisher2014ghosts} (\citeyear{fisher2014ghosts}) , since by definition a haunting is evidence of the presence of a thing not there.  A haunting exists entirely in our reaction to it - to the resonance between our environments and our psychologies.  The touch of the occult we see in psychogeography is found by engaging with those resonances.    

We can play an active role in these proceedings too - we are not necessarily the passive receptacles of signals from the environment. 
 It is possible to tune a sense of location and place, using the dial of our own personal experience. In this, a psychogeographer is less a diagnostic instrument in the style of Debord and more like a radio tuned to the unique psychological resonance that exists between you and the world.   Sinclair (\citeyear{sinclair2003london})  talks about the unique texture particular combinations of architecture and material bring to the sounds of the ring road.  This gives different parts of what might be seen as largely featureless concrete a personality that only exists in the intersection of experience.  Those experiences are personal, unique to ourselves.  The smell of a musty book means different things to different people - a grandfather's study, or a trip to the library, or hated classrooms.  These associations can act as an overlay, laminating the world in our own sense of our personal history.  All hauntings are true, because they exist in our minds - the evidence of things not seen and the intangibility of things only dimly recalled. 

Common to the literary, pseudo-scientific and explicitly subjective approaches of psychogeography outlined above is a desire to re-experience the commonplace.  This to extract from it such insights that can lend new richness to that which has become trivial or mundane (such as the M25 that encircles London).  In this, the discipline draws heavily - even if it is rarely formally acknowledged - from the works of the Russian formalists, particularly Viktor Shklovsky (\citeyear{shklovsky1917art}).  Communicating the insights of a psychological exploration often employs a principle known as defamiliarisation - the presenting of common experiences in an unfamiliar context so as to create new associations and to permit fresh perspectives.  Novels represent a kind of psychogeographical snapshot - or perhaps, less kindly, a form of stereotype or parody.   The resonance of the works of Ballard, for example, is explicitly encoded in a thematic and narrative context.  It has authenticity, but at a fidelty that is entirely negotiable at the whim of the author.  High-Rise as an example makes heavy use of social realism themes \cite{groes2012texture}, but the segmenting of the titular high-rise into social strata is a story-telling device.  Its artificiality undercuts the environmental truth that psychogeography attempts to capture.  That the book says something meaningful is not in dispute.   That it feels authentic is evidence of the author's own insights, which in turn are informed by authentic experience.  Psychogeographical novels though represent the field's equivalent of a 'strawman argument' - something explicitly constructed in the service of its message.  If it would have been more convenient for the high-rise to be a train  then the psychogeographical aspects would shift in service to the narrative.  

\section*{Principles of a Psychogeography of Imaginary Places}

As compelling as the practices of psychogeography in the real world are, they suffer limitations.  One cannot inhabit and experience the psychogeographical phantasms of literature - they are as much projections as they are reflections.  The pseudo-scientific psychogeography of the Situationists lacks rigour even in its expressed objectivity - it is impossible to separate the observer from the observed, and the derive does not incorporate the principles of ethnography and auto-ethnography that would control for such contamination.  The modern, empathic form of psychogeography is increasingly being forced to the peripheries as the banalification of urban spaces erases distinctiveness and character. This has the effect of turning the discipline into one increasingly confined to liminal and transitory spaces and the impoverishment implied by this relegation.  One cannot defamiliarise that with which the majority are not already familiar, and yet it is the familiar that is increasingly least fruitful when it comes to meaningful analysis.  The familiar has become too familiar.

We can then consider the extension of the principles of psychogeography into less tangible environments - those imaginary places that are represented by works of fiction, or places of the mind.   Aspects of psychogeography are a part of many creative disciplines - we have already briefly outlined the contributions of the literary psychogeographers to the philosophical heart of the discipline, for example.  Consideration and reflection with regards to the \textit{Mise-en-scène} of a stage, movie or television production is an established part of the storyboarding, stage direction and cinematography of an audio-visual production.   Much as with psychogeography itself, a formal, scoped definition of the \textit{Mise-en-scène} is elusive and ephemeral, indicating perhaps that the two concepts occupy similar conceptual real-estate.  It is reasonable to say though that television, cinema and the theatre are already comfortable wrestling with the amorphous complexity of psychogeography, in their own specific contexts.  

However, there are other fictive media in which psychogeography is not well established as a nexus between psychology and environment.  The concept of environmental storytelling is well established in gaming as a discipline, but almost always within a narrative, functional context.  And yet, video games represent perhaps the ideal medium for the exploration of a psychogeography of imaginary places.  They are literary constructions which we can inhabit and explore.  They offer opportunities to rigorously report on an environment without contaminating the insights through the simple toxin of presence - at least within single player experiences.  They are most often rejections of banality and the everyday, permitting a route for the peripatetic philosophy of the modern era to engage with environments that are at once familiar and yet utterly compellingly new.  Games are a construction of human imagination, at least for now (a point we will return to later).  They still offer opportunities to learn more about people and their relationship to places by virtue of being anchored in the common waters of human imagination.  In fact, a precedent already exists for this concept - that which is known in the psychogeographical literature as a Robinson \cite{coverley2018psychogeography} - a traveler of the mind, occasionally losing themselves in the amorphous space between that which is experienced and that which is imagined, as if there is no difference at all.  

The trajectory of modern psychogeography trends towards despair - as we discussed above, urban centres are subject to a deep and almost unyielding banalification that renders new experiences as remixes of old experiences.  How fitting then that video games offer an antidote to the commodification of modern life - the McDonaldization \cite{ritzer2021mcdonaldization} of experience, where all nuance and distinctiveness is flattened into that which is easily co-opted into viral semiotics and optimised functionality.  Games are resplendent in uniqueness and spectacle.  It is certainly true that we still see design convergence - after all, innovative design tends to simply become 'design' over time.  Uniqueness though remains something prized by players and something seen as intrinsically desirable by designers.  The construction of detailed worlds and stories - sometimes known informally as lore building - remains a vibrant part of the process.  All of this argues for the environments of video games to be treated as an almost ideal playground for psychogeographic practice.

The New Games Journalism \cite{golding2018writing}, popularised and then buried by its creatorr Kieron Gillen, was a brief-lived attempt to bring the subjective principles of the 'New Journalism' within the emerging conventions of contemporary games media.  In Gillen's manifesto, he outlined a celebration of the subjective and the personal - of an exploration of the individual feelings that a game might bring about.   The movement briefly sparked off some genuinely wonderful discourse, but soon imploded as it became seen as an enabler of self-indulgent melodrama, only barely respectful of the medium of games.  However, some of the most eloquent writers of games media emerged from the ashes of the movement, those who managed to use their engagement with play as a springboard to deeper insights.  At the core of the New Games Journalism manifesto is Gillen's evocative philosophical mission statement as to the job of a games journalist:

“This makes us Travel Journalists to Imaginary places. Our job is to describe what it’s like to visit a place that doesn’t exist outside of the gamer’s head […] Go to a place, report on its cultures, foibles, distractions and bring it back to entertain”

Indeed, it is this idea that is core to the earlier exploratory work the authors have done on the idea of video game screenshots as a form of travelogue for imaginary places \cite{gamboa2023screenshots}.  However, within this conceptualisation is the idea of a new game journalist as a reporter - someone packaging up insights for an external audience, with all the translation complications implied by turning the personal into the relatable.   As an antecedent to the idea of a of 'New Games Psychogeography' though it is a compelling statement of intent that captures much of the romanticism associated with the Baudelarian flâneur.   Here though we argue for psychogeography not necessarily as a tool for capturing communicable insight - although that is certainly one of the roles it can perform - but rather as a mechanism for players to meaningfully interrogate the parameters of their own unique experiences in reference to the things in a game that trigger psychological associations.  Sometimes those are elements of a game, of a game environment, or sometimes the environment within which a game was experienced.  

In this respect, the Psychogeography of Games has some connection with the Welsh concept of \textit{hiraeth} \cite{morgan2003marketing}.  Much like many components of the fragile conceptual architecture of psychogeography, this is a term that is difficult to precisely define.  Generally though it is most often translated into a longing for something that is gone, or departed.   Most often applied to a nostalgia for a Wales that may never have really existed, it is occasionally generalised into a sense of homesickness for a place that one has never visited and perhaps can never visit.  As the first author of this paper might say,  'I have been to Arrakis, I have fought on spice-tinged battlegrounds, and I feel melancholy that I can never go back'.  

It's possible to assemble the trappings of the experience.  I am in my childhood bedroom.  An Amiga 500 is on the desk in front of me.  Westwood Games' Dune 2 is on the screen.  A boombox is on the windowsill, in which I switch repeatedly between two albums.  These are  Jewel and Feeler , both by Marcella Detroit.  The lights are off.  The only illumination comes from the monitor, from which spill battles between the Harkonnen and the Atriedes.  I am desperately trying to mine enough spice to buy sufficient air support to destroy the launch sites of the devastating Death Hand missiles that my enemy is raining upon my barracks.  The sounds, the scenes, the sense of urgency - all come together to create a snapshot of my youth that exists only in the intersection of all those parts. Every part of the experience can be reconstructed, at least in principle, except the naive child who wasn't yet carrying the weight of adult responsibilities.  The psychogeography of gaming, like the concept of hireath, is inherently marinated in melancholy.   Paradoxically, for a medium in which worlds are more often static than they are dynamic, these are worlds characterized by transience.   We engage with a nostalgia for a thing that is gone, and can't ever come back.

In an earlier section of this paper, we discussed the idea of psychogeography as being particularly potent within liminal spaces, and how the occult is often concentrated within the transitory.   Here we posit the video game as a kind of modern faerie circle - a point of entry whereby we take leave of reality and temporarily enter another.  Much in the style of a medieval traveler who unwarily takes a nap in the forest only to find themselves transformed or transported as a temporary denizen of a fantastical world.  We have long discussed the idea of certain pursuits existing within a kind of magic circle \cite{heron2014you}, or rather a discontinuous moral space where the rules of the external world do not apply.  Within a boxing ring, it is appropriate for the fighters to punch each other repeatedly for the entertainment of the audience.  In a supermarket, such actions would receive censure and prohibition - people would intervene, and be seen as upstanding for their efforts.  If one rushed a boxing ring to break up the fighters, that person would be considered the party at fault.  This conception of the magic circle is a useful theoretical tool, but perhaps we have not gone quite far enough.  The faerie circle concept is one that acknowledges the act of transfer itself - the movement between this world and another - where our perception inhabits a space that is not only morally discontinuous but also perceptually so.  'Life is the sum of all your choices', which is a quote often attributed to Albert Camus, which has been further expanded upon to argue that our life is the sum total of the things to which we pay attention.   In that regard, if we choose to pay attention to what is happening within a video game, why should that not receive the same level of import as that which happens in the physical world around us? 

After all, within a video game we tend to find ourselves centred.  No longer an observer, but rather the prime mover.  We are not the Debordian Situationists, walking into unfamiliar contexts and drawing lofty conclusions from erratic sampling.  What life exists around us in a game is simulated for our pleasure.  That which we experience is true, albeit sometimes only in a multiple universes perspective \cite {heron2017pacman}.  What we subjectively experience is fact, for us.  There is no true canon of what happens in a game, but there is a personal canon that is uniquely applicable to us.  A canon that carries with it a resonance and emotional energy that is specifically attuned to our psychology and the specific moment in time in which we experience it.  

With regards to the physical constraints of the real world, we return to the idea of the placement of a harbour - a harbour exists by water because it would be ridiculous for it to be built in the middle of a land-locked city centre.  Our physical universe creates a set of principles which guide all other things that follow.  It is not possible within a game though to convincingly argue 'it just is', since the physical reality of a video game is intrinsically additive.  What is implemented, and what is not implemented, is a product of developers and publishers.   This concretizes politics within a game in a way that cannot be hand-waved away with the counterbalance of time.  There is no, 'According to the land rights granted in 1872, we need to demolish your house in 2024'.  The harbour of a video game can exist anywhere, because the river that it depends on can \textbf{flow} anywhere.  Games are full of geographical and algorithmic decisions of the kind that geology does not make.  Water flows where it flows.  Video game rivers take the shape the developers impose.  As such, a close reading of a player's engagement with their virtual environment offers a route to meaningful annotation regarding the subtle interaction of intent, implication and experience.  It is sometimes said that all games are political, and it is this feature that is most responsible for that.  Everything within a game is intentional or endorsed, and such decisions are inherently political even when they may not be explicitly ideological.

We might in fact think of this suggestion - that of a Psychogeography of Games - as an almost full circle return to the origin of the discipline, which began in the works of writers before it was forcefully transplanted into the real world by Debord and the Situationists.   In any case, it is the opinion of the authors that this form of mindful engagement with the psychogeography of gaming offers a lens that can generate new insights, in new forms, that offer both personal illumination and actionaable design insights.  However, this is not a discipline that can be translated entirely into the explorable context of video game environments, because the nature of gaming is that - as viewed through the faerie circle - it does not bear a one to one fidelity to the physical assumptions that underlie psychogeograpgical practice.  Time, and space, and geometry are not so much non-Euclidian as they are a bespoke invention tailored to the needs of each gaming experience.

We argue here too for some urgency in the transplant of this practice.  To paraphrase Kurt Vonnegut, If banalisation is the lead in the water that drives psychogeographers mad then we must get ahead of the lead that is coming down the pipeline into gaming experiences.   What is generative AI, after all, but an algorithm for banalification?  What else but a mechanism for recycling counterfeit creativity, passing off soulless facsimiles as new experiences?   Generative AI is already beginning to find safe harbour in numerous disciplines of game design - environmental construction; dialog systems; level design; and more.  If one is interested in interrogating the soul of a video game through a psychogeographical lens, one must do so before the soul is replaced with the outputs of silicon.

\section*{Practical Considerations for a Psychogeography of Video Games}

To fully transplate this discipline into gaming, it's necessary to examine the underlying assumptions of psychogeography that do not hold true in gaming environments.  Chiefly these relate to the nature of time, space, and the underlying physics of reality.  Within video games these are situational, malleable, and suspend-able.  Relationships may become elastic, flexing and unflexing contextually.   We argue though that this is not a limitation of the approach, but rather a new vector of analysis that offers exciting opportunities for insight.  We do not need to query the physical world for what the geology of an real-world environment is saying.  It simply is.  Absent the presence of certain pharmaceutical interventions, reality has a one to one scale with our physical world.  Our perceptions may shift, but the physical reality does not.  For one to travel from A to Z, one must traverse all points in-between.  None of this is true of video games as a whole, and even the rules established within any given video game may be negotiable.  These offer a set of interesting analytical lenses that a gaming psychogeographer can contemplate when it comes to using the technique within play experiences.  In this respect, we can consider the Psychogeography of Video Games as an extension of the literary tradition of a close-reading - heavily drawing in the aspects of embodiment that other more passive forms of media do not encode within the text.

\subsection*{Time}

When we consider the impact of time within a video game, we must contemplate the relationship between in-game time and the external time we experience - or, time as our player-character sees it versus time as we do.  It is rare that a game with day/night cycles affixes these to a twenty-four hour clock, even if the time represented within a game may do so.  Time usually runs quicker in a video game - in Red Dead Redemption 2 (RDR2)  an almost perfect example of a game that attempts to enforce fidelity to the real world in every conceivable dimension - two seconds of real time represents a minute of game time.  A full day in RDR2 lasts around forty-eight minutes of real, external time.  Moreso, time has an elastic property that allows sections to be skipped (through sleeping, for example) or sped through.  The relationship between our time, and the game time, is not uniform.  Animal Crossing as a counter-example has a game clock that is pegged to the time-zone of its player.  If you log into the game in the evening, your island will likewise be in the evening.  Shops have opening hours, and a player working anti-social hours in the real world may find themselves raging against the lack of 24hr supermarkets.  Time cannot be accelerated, or slowed down, although canny Switch users did find out it was possible to 'time travel' by changing the time of their device's real world clock.  Dungeons and Dragons  and the various games built upon its tabletop ruleset, slices combat up into sequential turns of six-seconds duration, each of which may take many minutes of discussion, planning, dice-rolling and looking up of rule interactions.  A single combat encounter may take thirty seconds in game-time and yet last an hour of real-time.   Traversing vast distances, on the other hand, may be handled 'vignette style' in which a journey of a thousand miles may be resolved as a series of individual encounters, combat or otherwise, that turn a journey of two game weeks into five real world minutes.  Vampire the Masquerade as a counter example employs a completely malleable concept of time, encoding the convention that 'things take as long as they take'.  

The various civilization strategy games offer an even starker derangement of time, and consequently of geography.  To begin with, each turn may cover 40 years of ingame time, making the construction of a settler or the raising of a militia unit a task that takes several in-game centuries.  Many units can move a single map square at a time, and so moving a unit of warriors from one side of a small island to another may also take hundreds of years.  Technological development is similarly slow, as are the wheels of trade and diplomacy.  As the game progresses, the scale changes - from 40 years to 25, and then to 20, and eventually to a year, or less, per turn.  What this implies about the nature of physics is outside the scope of the game's mechanics, although ideologically it is designed to simulate the accelerating pace of progress of our own planet.  The anachronisms that are produced by this odd system of chronology are manifest, including cavalry units that become quicker with time, and wars fought between bronze-age spearmen and modern armour units.

Many games offer a pause facility which allows for the consensual suspension of the passage of time.  Pausing might also be incorporated into non-diegetic activities - those explicitly 'out of game' activities that serve as the backbone of bureaucracy in play.  Inventory management, menu navigation, and more all tend to result in an automatic pause.  This can be explicitly immersion breaking... Skyrim for example allows one to pause a frantic combat by opening the inventory screen, from which one might guzzle down a dozen healing potions free from the consequences one would associate with that activity in the real world.  Other games continue to allow time to pass when within menu screens, creating a potential tension when one must consider inventory management as a form of inattention - a window of vulnerability in which one might find oneself attacked or disadvantaged.   The way in which a game suspends the passage of time is, like much of game design, contextual and often in service to greater design goals.   The consequences of the mechanisms of pausing and unpausing offer a meaningful topic for meditation.

Absent some exotic relativity effects, psychogeographers in the physical world experience time at approximately one second every second.  One need not interrogate what the scale of time is telling us about the environment, but assessing this within a video game can say a lot about the nature of play and design.  Particularly when it comes to the pacing of an experience and how the player's centralised reality must be synchronized with the suspension of disbelief required for the simulation of the rest of the game.

\subsection*{Space}

Video games, particularly open world video games, often pride themselves on the size of their maps.  However, measuring the pace of an avatar versus the scale of an environment can reveal vast landscapes are rendered almost comically small through the application of arithmetic.  To traverse the entire map of Red Dead Redemption 2, on foot, takes approximately three hours, according to the YouTube channel Same Tie Studios who engaged in an almost psychogeographic experiment on the cartography of the game.   Arthur Morgan, the main character, is in his mid-forties given the game's timeline (born 1863, died 1907) which would give him an average top-end walking speed of perhaps 3.2 miles per hour.   Walking one end to another then gives an approximate size of the game map of 9.6 miles.  However, that is only true if we take time at a 1-to-1 scale, which we have already addressed.  The relationship between perceived and real space is elastic, but the relationship between time and space too is difficult to pin down.  One cannot simply focus on how big a game is, but rather on how big a game \textbf{feels}.  Red Dead Redemption 2, at least in the opinion of the authors, is a game that feels geographically unconstrained in a way that the actual dimensions of the map do not reflect.

But even this is only part of the psychogeographical puzzle of space in a game.  Game mechanisms, common to many games, frustrate a simplistic interrogation of the environment.  Will Self \cite{self2007psychogeography} , discussing his psychogeography via vehicles, has sometimes described the approach as a kind of 'jump cut', where the passive experience of being on a train or a plane is a real-world star-wipe.  In video games, the parallels for this would be the loading screen, which obscures the continuity of environments, or perhaps fast-travel.  Loading screens are commonly used to transition players from one map to another, such as moving from an over-world map into an abandoned building or leaving one city to go to another.  It is not necessary in a game, in other words, to traverse all the intervening points to get from A to Z.  The character can simply be moved, or the world deconstructed and reconstructed around them.   To go with the derangement of time implied by the design of a video game, we must add a derangement of place.   Cut-scenes too - those occasional cinematic interventions - may take the pacing of this derangement of place out of our hands.  A cut-scene may take us from the interior of a building to the exterior roof of another city, in service of the narrative, without us having moved the character anywhere.  In fact, cut-scenes don't just create a derangement of spatiality, they also fundamentally break the thread of perception a player may experience.  Nowhere is that more in evidence than when one mistakenly, or intentionally, skips a cut-scene to find themselves suddenly facing down hostile enemies in unfamiliar terrain.  An eerie sense of displacement is the common psychological consequence.

In reaction to this, we can see proto-psychogeographical approaches taken by many players.  A challenge players of Skyrim occasionally set themselves is to traverse the landscape without using any of the fast-travel points permitted.  If one must deliver a letter from Jarl Balgruuf to Jarl Igmud, one must walk from Whiterun to The Reach.  This is in contrast to the game mechanisms which allow a player to teleport from one place to certain other previously discovered locations.  Fast-travel removes the tedium from the journey, but also the risk and the opportunities to explore hidden ruins and discover forgotten secrets.   It also obscures the scale of a world, and those players dependent on its functionality will often find themselves adrift should it become unavailable or stop working.  Fast travel in many respects creates a conception of a world that mirrors that of how children understand their own geographical context.  a child can certainly identify 'my house', 'my school', 'the houses of my grandparents' and so on.  However, if they are most often carried or driven to these places, that same child may not be able to articulate the context between them - how far away they are, in what direction, and what roads must be walked to get there.  A visitor to London who has only explored the city via the subway can almost certain relate to this.

The fragmentation of geometry that games permit creates equally fragmented mental associations in players.  Some of this is down to player convenience.  Some of it is down to fundamental technological limitations - squeezing through tight passageways, or elevators, are common ways to maintain the fiction of an open environment without anything as stark as a loading screen.  The overall effect however remains the same - disjointed perceptions of time and space that are decoupled from the physics of a consistent universe.

As has been remarked in earlier sections, video game construction is an inherently selective additive process - the mechanisms of reality - or unreality - that are incorporated must be selected and implemented.  One of the mechanisms that is open for consideration in a video game then is the relationship of the player with regards to modification of the environment - the graph in psychogeographer.  A physical world psychogeographer, walking in wet soil, will leave footprints.  Their scent may linger.  Their sounds will echo.  There is a physical annotation of our environments that occurs as a simple consequence of our presence.  Red Dead Redemption 2 was noted upon release for the comprehensiveness of how it allows a player to leave trace evidence of their passage.  One will leave tracks in the snow, hoof-prints in the dust, and trails in the mud.  When one drags a victim behind a horse, there is unmistakable trace evidence left behind.  Similarly, clothing can acquire annotations linked to the environment one has explored - shirts can become bullet-holed and bloody, boots can accumulate dust and dirt.  The extent to which a player character influences the environment, and the environment impacts the player, can be a useful focus for analysis.  

Indeed the principles of a dérive themselves were originally considered, by the Situationists, to be a form of physical automatic writing.  In this frame, we can consider the outputs of our engagement with an environment - physical or otherwise - as an informal output of the process of psychogeography.  We might see this as a form of the game being in dialog with itself through the player as an interlocuter.  Capturing the essence of what that conversation is about can permit a deeper understanding of what the game is saying.  The Surrealists, one of the early influences of the Situationist movement, employed the term \textit{deambulation} in which the goal was 'the achievement of a state of hypnosis by walking', and 'it is a medium through which to enter into contact with the unconscious part of the territory' \cite{bassett2004walking}.  The attainment of flow is an oft-stated goal of gamers, and much game design is focused on permitting the circumstances under which flow can emerge \cite{chen2007flow}.  A form of deambulation is almost intrinsic to attainment of flow.  Trace environmental evidence, like the pen of a medium in an occult trance, is the evidence of things not consciously recorded.  Games offer an opportunity for the culmination of surrealist intention - the resolution of the states of dream and the real world into an absolute blended super reality, or what was sometimes termed a surreality \cite{breton1969manifestoes}.   The often inconsistent spatiality of game environments can underpin this.

Nowhere is this more evident perhaps than in those games that are delivered through virtual reality - a subset of gaming that perhaps deserves an analytical deconstruction of psychogeography in its own rights.  However, the dream represented by a virtual reality game, constrained by the physical environment within which it is experienced, can create a strong sense of disassociation.  One occupies a liminal space that does not truly honour any of its implied contracts.  When playing Superhot in VR, for example, the first author of this paper took refuge behind a table in the game world.  Protected from bullets, he then stood up, leaned on the virtual table to acquire a targeting solution, and fell flat on his face in the real world.  The viral footage of snooker champion Ronnie O'Sullivan falling over after attempting to position himself across a VR snooker table is compelling evidence of the derangement of spatiality that video games employ.  Few might be quite so effective at ferrying a player through the faerie circle to a genuinely new perception of the external world as virtual reality.

\subsection*{Physics}

The nature of an environment in a video game is a function of a design, but it's also a function of technology.  We might think of the technical architecture of a video game as offering a framework for assessing its physical properties, and the implications of the physics it embodies.   Many game engines exist, and are designed towards different goals.  Engines such as Unity, Godot or Unreal are general purpose engines, permitting a common framework for many games to be created in an agnostic manner.  Other engines are particularly tuned towards other goals, such as Inform, Inkle, and Twine which are primarily used for the construction of text games.  Encoded into each engine is a set of assumptions regarding how games should be made, which processes should be supported, and the weight to be given to particular subsystems.  As such, individual engines have a kind of 'aura' which can be perceived - by the informed observant - through the way a game built upon an engine presents itself.  Certain quirks or affordances become identifiable.  At the same time, the emergence of middleware can smooth away incompatibilities or inconsistencies in behaviour.  We tend though to see certain kinds of games most prominently delivered on certain kinds of engines.  Large, open-world games where fluidity of the environment is to be stressed will find themselves best served by something like Apex  which can support huge maps served on-the-fly through streaming.  Expansive spaces, immersive terrains, vast vistas, and uninterrupted traversal are the shibboleths of Apex.  On the other hand, Bethesda's Creation engine  relies upon transition zones where large labyrinths and buildings must be accessed via loading screens.  While these secondary locations are likely richer and more detailed than what can be supported seamlessly by Apex, they are also accessible only through a system punctuated by immersion-breaking non-diegetic elements which lead to claustrophobic, spatially disconnected spaces.

The quirks encoded into game engines, or sometimes into the specific employment of a game engine within a specific game, also introduce an aspect of game psychogeography that has no antecedents in the real world - that of the existence of glitches, exploits and bugs.  That said, given widespread belief in the Mandela Effect and certain principles of occultism and mysticism, perhaps we should concede that this point is at least somewhat arguable.  Suffice to say though that the engines that run games do not simulate things perfectly, and computational constraints are often employed to ensure that a 'good enough' performance can be obtained at speed.  The exact mechanisms used for collision detection, for example, can create constraints on gameplay.  One common example of this is known as 'tunneling' - when objects that should collide instead move through each other.  Simple collision detection is a frame-by-frame comparison of co-ordinates and dimensions, and collisions are triggered when one object occupies the physical space of another.  Movement is traditionally handled by translations of co-ordinates.  Collisions are calculated within frame, usually after the various translations are performed.  However, if an object never occupies the space of another, it will never trigger a collision - fast moving objects might move 'through' another without ever occupying its bounds.   To solve this, more complex collision detection routines may be used - at the cost of computational time and system complexity - but no collision detection system works correctly in all circumstances.

Knowledge of this underlying truth of engine architecture creates opportunities for exploits.  One might throw their character against a terrain wall, for example, attempting to find the combination of velocity, angle of approach and timing that tickles a weakness in the collision detection.  Diligent probing of vulnerabilities (such as when multiple objects work together to create barriers, as might occur in the right-angle intersection of two segments of walls) can permit access to parts of the game that a developer did not intend players to see.  Many colliders in a video game are two-dimensional - they have width and breadth but no depth - particularly those related to terrain.  This is to say - once you get passed the collider check, the game won't check again whether you're supposed to be where you are.  It is possible to peek behind the curtain of a game environment, and for one to bend its rules in favour of unsupported ends.  We are not always required, within a game, to accept its physical laws as inviolable.

The way a game presents spatiality has already been discussed, but relevant too is the nature of the way the game enables players to traverse those spaces.  The forms of movement permitted by the underlying physics of a game can reflect the priorities and properties of our characters, which in turn can influence the extent to which psychogeographical techniques can be applied - or rather, the flavour of psychogeography which is most applicable.  Swift, unfettered movement can trigger the same absent-mindedness one would experience while driving a car.  Webslinging in the recent Spiderman games is largely frictionless and soothingly rhythmic, which means one might traverse miles of in-game territory without ever truly taking anything in.   One can see a similar thing at hand in Saints Row 4, in which the main character is granted super-powers that permit for vast leaps and the ability to effectively fly, turning terrain traversal into a task of 'avoid the most prominent obstacles'.  Other games  emphasize the slow, physical movement of a real body such as the family of games sometimes derisively labeled as 'walking simulators', which the authors of this paper have - largely unsuccessfully - attempt to rebrand as 'empathic puzzlers' \cite{heron2015all}.  These games permit,and indeed encourage making space, for the time required to meditate upon one's surroundings in the manner we would associate with physical psychogeography.  We discussed already the idea of fast travel, but the physical properties of movement are just as impactful - to what extent are we pedestrians, mindfully exploring the world around us, versus anthropomorphized automobiles, speeding to the next destination.  Psychogeography is a process in which the journey subordinates the destination, and the relationship between our characters and the traversal of the environment is a powerful moderator of that.  Some games are better than others when it comes to engaging with our embodied presence in an environment.

\subsection*{The Inconsistency of Experience}

We observe the world through faliable biological mechanisms - aging eyes, unreliable ears, the erratic and inconsistent models off our own perception.  However, we generally assume - as a pragmatic convention if not necessarily a philosophical truth - that there is an underlying reality that we are perceiving.  Short-sightedness may impact on how that perception manifests, but we do not generally believe our blurry vision reflects a deeper truth of the world.   Again, this convention does not hold in the world of video games where technical considerations can change the underlying representation - and even behaviour - of the worlds through which we wander. Processing power can impact on the performance of physics systems.  Grapbical architecture can change not just the resolution of in-game artefacts but even their presence.  Lower-end graphics card will simply not be able to execute upon aesthetic techniques that may impact greatly upon the psychogeographical experience.   There exist platform difference too, both in terms of how games are implemented but also in the wider ecosystem of how they feel to play.   PC gamers used to an Xbox controller may find their experience on a Nintendo Switch frustrating due to the fact the A button is located in a different place - a joyful experience may become frictive and irritating through a kind of morse code of muscle memory.

One may find themslves lost in meditation staring into the reflected neon cityscape of Night City as it is ray-traced into an oily puddle on a busy road only to find that another player, with less sophisticated hardware, sees only a blurry impression like a low-resolution pointilist sketch.   Within a video game we may not share the same story \cite{heron2017pacman} but also we may not even share the physical realities of the environment and the consequences of our actions.  We need to mindful, in other words, that the experience of a video game is quantum - a kind of multi-worlds interpretation of play - and our reflections on our experiences must be contextualised within that frame.  Everything from the graphical settings (low, medium, high, ultra-high) to the resolution of the screen, to the frame rate, to the degree of camera shake, film grain and more all subtly and not-so-subtly change the experience of play.  A psychogeography of imaginary places - particularly those rendered on technological platforms - cannot be agnostic of the consequences of that technology.  Again, such complexity does not in the view of the authors represent a problem but rather an interesting topic of analysis - to what extent do the capabilities of our hardware platforms change the psychogeographical landscape of the games we play?

\subsection*{The Secret Alphabet of Play}

Iain Sinclair, a noted psychogeographer who considers the analysis of the city of London to be his life's great project, has occasionally described his work as a kind of 'ambulant sign-making' \cite{sinclair2003lights}.  Work where walking is a process of stitching together disparate and disconnected urban locations by forging a narrative and emphasising connection where none naturally exists.  In his conception, the psychogeographer is an analyst discovering hidden commonalities and subtle relationships.  Psychogeography here is an active job of identifying latent meaning and bringing it to the fore.  However, Sinclair also acknowledges that such stitching will always be personal, difficult to articulate, and cryptic in its implications.  The physical movements of the characters across their territory might spell out the letters of a secret alphabet.

This concept underpins much of the psychogeography of play.  Where Sinclair has his roads, byways and landmarks a gamer has routes, runs and grinds.  Consider here the design of the Soulsborne games, particular the Dark Souls series and Elden Ring.  More than any other series of games, we would argue that these are perhaps the most explicitly (if not intentionally) psychogeographical games that exist.  These are a combination of well balanced game mechanics that demand mindful engagement with a resonant environment. 
 The
 storytelling is articulately expressed through the geography and visual design of a world that is comprehensible to an inquisitive mind.  The experience of Dark Souls is one that cannot be easily generalised, the relationship that a player has to their experience is a unique tale of personal triumphs and failures.   

 The repeated journeys from bonfire to boss in Dark Souls can be thought of as the carving of a psychic glyph in the landscape, Sinclair style.  The meaning of that glyph - encoded in a player's psychology - is cryptic.  Secret.  Repeatedly traversing these routes, deviating occasionally based on aggregated psychic damage, accumulates a tenor that is shaded by the specific player's experience.  Uncertainty becomes fear, becomes determination, and becomes despair and despondency.  The boss seems so difficult, the player character so ill-equipped to deal with it.  Early composure becomes a sense of futility.  And then, after a while - perhaps it's attempt five, perhaps it's attempt fifty, the player does better than they expected.  Maybe it was a new strategy, maybe it was just that they started to jive with the timing of the battle.  Despondency becomes hope.  Hope becomes determination.  They get the boss down to fifty percent health.  Then forty.  Then thirty.  Hope becomes excitement, excitement becomes anticipation.  Twenty.  Ten.  And then that anticipation becomes triumph as the boss is slain, the challenge surmounted, and the route completed.  A secret alphabet in a secret text, held only in the player's emotional relationship to their environment.  They'll pass by this way again in the future and feel the letter pulsing.  

 Perhaps here it would be more appropriate to think not so much of a secret alphabet but rather a secret melody.  If one could track, pheromone style, the accumulation of steps in the game map - cross-referenced against psychic resonance - it would play out the history of an experience in a series of individualised, psychic runes.  If the maps of every player were merged, shared runes would begin to materialise from the chaos - arcs of experience that, while unique, create commonalities.  These three-dimensional glyphs of time, space and emotion would encode shared, and diverse, experiences.  The low minor-chords of tragedy gradually becoming the triumphant chords of catharsis.   The landscapes of gaming gradually become emotionally resonant sheet music.  The suggestion of a secret alphabet implies something like a sorceror's hidden grimoire.  Perhaps the more appropriate comparison within games is that these experiencs should instead be encoded in a kind of emotional hymnal, where the rhymes and rhythm are embedded deep within our own synapes.

All of this is to say that the experience of gaming - its loops, its pacing, its expectations of play - all fall within the framework of psychogeography.

\subsection*{Psychogeographical Anti-Patterns}

Not all games are going to be amenable to psychogeography.  This is less to do with specific genre and style of game and more to do with how certain gaming conventions are emphasized or de-emphasized within a design.  One might consider for example the presence of in-game maps replete with quest markers and points of interest to be almost wholly counter to many of the principles underlying the technique.  If a dérive is, explicitly, an unplanned excursion through a landscape - it is hard to imagine how its spirit can thrive when one is constantly being directed to a destination.  Such approaches are all geography and no psychology - less a meaningful meander through a world and more a frantic dash to complete a to-do list of obligations.  It is not necessarily the presence of maps, audio logs, and other collectible ephemera that represents an anti-pattern, but rather the intangibility interwoven with the obtrusiveness of its presence.  Hand-made maps, abstracted physical feelies, and other accouterments represent a greater degree of distance between the representation and fidelity to the source.  On-screen markers, information laden HUDs, in-game navigational cues - all of these are incredibly useful to the gamer who wants to get things done.  Simply existing in a place and permitting whim to direct your feet cannot happen in an environment with constant exhortation to productivity.  The genius loci of a game must be compatible with a restless spirit and a dash of wanderlust.

Here we might draw in an additional mote of mysticism.  In Ancient Greece, an Omphalos was a religious stone marker that indicated the spiritual centre of a thing.  One might argue that the Omphalos of a game is to be found in relation to where our focus is most often drawn.   This locus of attention is the thing that dominates our perception of a game.  Do we spend our time with our eyes affixed to a mini-map so we know every marker around us?  Do we focus on the main display of the game only long enough for us to make sure we're not walking into a wall?  It's hard to shake the feeling that some video games are the equivalent of staring at our phones while walking through the busy streets of a city.  Others eschew adornments except for those which are mandatory, and relegate them to corners of the screen.  

Other games are more subtle about their cues, but are still manipulating the environment to direct the attention of the player towards points of particular importance - the injection of attentional mirages into what may be perceived as an entirely unstructured experience.  Players may be directed to particular locations through subtle cues in lighting, or signs of activity, or other such unobtrusive, diegetic signposts.  Nowhere is it as blunt an instrument as a quest marker, but more like a kind of subliminal messaging in the layout of levels.  The real world is obviously full of enticing distractions, but rarely put to a coherent purpose in the way that a level designer will use brightness and shadow to covertly map out the path to a destination.  In a video game, one might be following artificially designed desire paths of which  they aren't even aware.   

Core too to the experience of psychogeography is control over attention, and this is something that is sometimes not within the player's domain in a video game.  In a book, we get the description the author gives us.  In a movie, our eye goes where the camera directs it.  In the real world, we can walk backwards, look at right angles to our movement, stop and contemplate an environment.  While we can't do any of this without consideration of the consequences - stopping to contemplate some architecture in the middle of a busy road is not going to be a successful strategy for anyone - we can do it widely and freely.  A video game may not give us that same level of control over our character.  Pacing may be dictated by cut scenes, or by in-game footage intercut with Quicktime Events.  We may find ourselves disadvantaged if we choose to tarry.   We may be stuck with a camera that auto-follows at a particular distance and angle, and we may find that it doesn't properly function in tight constraints or in complex geographies.   Or, we may be given what seems like full control over a camera and the character only to find that control is conditional - cameras may gradually return to an angle chosen by the developer, or they may be afflicted with motion blur, or shake, or other 'cinematic' effects.  We may be able to configure the camera in the settings of the game, but we might not.  Games exist in an awkward space where the gamer has control over much of what happens within a game but the game developers may insert their own preferences in service of storytelling, ambiance, or simple bloody-mindedness.  

We may be permitted choices within a game, or we may have control over nothing but the pace at which we achieve game goals.  The environment may be amenable to exploration, or it may be essentially set dressing for a pre-determined adventure.  Psychogeography can still be conducted within these constraints - and indeed, within almost any constraints.  Consider the text 'A Journey Round My Room' \cite{de1899journey}, in which Xavier de Maistre conducts a psychogeographical investigation of the single room in which he was undergoing a form of house arrest.  The activities of a Robinson do not need the full freedom of the open city to flourish.  However, such approaches require a specialised sense of purpose and often lack the variety required to compel an external reader.  There are fertile and infertile grounds for the psychogeography of play.

\section* {Conclusion}

We do not claim in this paper that the psychogeography of play did not exist previously.  There are already abundant sources that show peripatetic players engaging in meaningful meandering through game worlds.  Walking from one end of Red Dead Redemption 2 to the other.  Traversing Skyrim without fast travel.   Meditations upon sound and embodiment and the freedom of the landscape in Legend of Zelda: Breath of the Wild.  There is much active participation around engaging with game environments for their own intrinsic value.  Similarly, we do not claim here that applicable techniques from other disciplines have not been profitably employed in the analysis of games.   Literary close-readings are uncommon, but exist \cite{bizzocchi2011well, tanenbaum2009close, bostan2022using}.  An appreciation of environmental storytelling is embedded in the discipline.  Open world design explicitly accommodates the goals of a dérive as a matter of best practice.  However, all of these approaches and considerations exist largely in isolation, intersecting only infrequently and without deeper cohesion.

We propose here that psychogeography, as ill-defined and fragmentary as it is as a discipline, is nonetheless a unifying framework that offers a profitable route to scholars of games who are interested in extracting meaning from the games they play.  The outputs of psychogeographers have offered up perspectives and nuances of the urban, suburban and rural landscapes for decades, and have produced numerous meditations upon the relationship of history to tradition to architecture.  Within video games, the discipline has some complexities in application that stem from the fact that we leave the easy assumptions of the physical world - that time, space and physics are at least somewhat reliable anchor points for our experience.  We instead move into environments where all of these things are contextual and elastic.  However, rather than this being a weakness of the approach we argue that this is one of the great strengths of the medium.  The complexity, nuance and richness of analysis permitted by a psychogeography of imaginary places is a most pleasing output as opposed to a disciplinary morbidity.

We offer this paper not as a comprehensive overview of psychogeography nor a particularly exhaustive overview of how it might be applied in gaming.  Instead we offer this as the start of what we hope to be a vibrant conversation on the role that psychogeography can play in exploring the contours of gaming, and our relationship with play and our virtual environments.

\footnotesize
\printbibliography

\end{document}